\documentclass{article}
\usepackage{graphicx}
\usepackage{dcolumn}
\usepackage{bm}
\usepackage{hyperref}
\usepackage{subcaption}
\usepackage{fancyhdr}
\usepackage[utf8x]{inputenc} 
\usepackage{authblk}
\usepackage{array}
\usepackage{multicol}
\usepackage{float}
\usepackage{wrapfig}
\title{Resist and Transfer Free Patterned CVD Graphene Growth on ALD $MoC_{x}$ Nano Layers}
\author[1]{Eldad Grady}
\author[2]{Chenhui Li}
\author[2]{Oded Raz}
\author[1]{W.M.M. Kessels}
\author[1]{Ageeth A. Bol}
\affil[]{Department of Applied Physics, Eindhoven University of Technology, Den Dolech 2, P.O. Box 513, 5600 MB Eindhoven, The Netherlands}
\affil[2]{Institute for Photonic Integration Eindhoven University of Technology Den Dolech 2, 5612 AZ Eindhoven, the Netherlands}
\affil[ ]{\textit {Corresponding author: gradyel@protonmail.com}}
\date{}
\begin{document}
\maketitle
\begin{abstract}
Multilayer graphene (MLG) films were grown by chemical vapour deposition (CVD) on molybdenum carbide ($MoC_{x}$) substrates. We fabricated the catalytic $MoC_{x}$ films by plasma enhanced atomic layer deposition (PEALD). The mechanism of graphene growth is studied and analysed for amorphous and crystalline $MoC_{x}$ films. In addition, the unique advantages of catalytic substrate PEALD are demonstrated in two approaches to graphene device fabrication. First, we present a complete bottom up, resist-free patterned graphene growth (GG) on pre-patterned $MoC_{x}$ PEALD performed at 50$^{\circ}C$. Selective CVD GG eliminates the need to pattern or transfer the graphene film to retain its pristine, as grown, qualities. Furthermore, we fabricated MLG directly on PEALD $MoC_{x}$ on 100 nm suspended SiN membrane. We characterise the MLG qualities using Raman spectroscopy, and analyse the samples by optical microscopy, scanning electron microscopy and X-ray diffraction measurements. The techniques of graphene device manufacturing demonstrated here pave the path for large scale production of graphene applications.

\end{abstract}
\fancyhead[]{Preprint}
\section{Introduction}
Next generation graphene-based applications would require a resist and transfer free graphene growth in order to realise the promise of graphene for flexible electronics\cite{strong2012patterning,kim2011stretchable}, and sub 10nm nodes' interconnects \cite{wang2017replacing}. Graphene based interconnects would significantly reduce the RC delay and thermal budget which is currently the bottle neck in sub 10nm transistor nodes \cite{politou2017evaluation,awano2009graphene} due to the excellent room temperature carrier mobility and thermal conductivity. However, graphene based devices exhibit far lower carrier mobility than the theoretical predictions promise. Encapsulation of graphene with lattice matching dielectric material such as hBN shows remarkable improvement, but performance is still subpar \cite{rakheja2013evaluation}. The main challenges of current commercial chemical vapour deposition (CVD) grown graphene are resist residue due to target transfer process \cite{kang2012graphene,choi2017influence}, and patterning of graphene, which degrade the graphene quality, resulting in the far lower carrier mobility than theoretical predictions. Known techniques include photolithography, ion beam milling \cite{lemme2009etching}, shadow masking \cite{kim2014direct}, area selective passivation layer \cite{hofmann2014scalable} and stencil mask \cite{yong2016rapid}. Despite these tremendous efforts, the goal of a scalable, low cost, high quality patterned graphene using commercially available tools, has not been achieved. Photolithography methods leave resist residue, and ion beam etching causes broad lateral damage close to pattern area due to scattering ions \cite{canccado2011quantifying,thissen2015effect}. Patterning graphene with oxygen plasma inevitably oxides graphene edges \cite{childres2011effect}, and the stencil mask poses resolution limitation . Furthermore, transfer of graphene from one substrate onto another degrades the graphene quality due to induced film compression and tension, and potential contamination between the graphene and target interfaces.
A wholesome solution that avoids both the need to pattern the graphene or to transfer a pre patterned graphene onto a target, has yet to be introduced. While advancements have been made toward transfer free graphene grown on patterned Mo catalytic layer \cite{vollebregt2016transfer}, resist residue on the underlying growth substrate cannot be fully removed. As defects, non uniformity and contamination will imminently translate to graphene defects, it is vital to achieve a resist free catalytic surface, in a uniform and conformal layer deposition. Atomic layer deposition (ALD) is a cyclic, broad temperature window soft deposition, with precise thickness control due to its self-limiting nature. It allows for unmatching conformal deposition on high-aspect ratio (HAR) objects. Recently, we demonstrated the plasma enhanced ALD (PEALD)  of $MoC_{x}$, with excellent composition control \cite{grady2019tailored}. Consequently, we presented the growth of multilayer graphene (MLG) films on these substrates, and the correlation between the catalytic substrate physical and chemical properties to the grown MLG \cite{grady2019control}. In this work, we describe the growth mechanism of graphene on $MoC_{x}$, and compare between amorphous and highly crystalline catalytic substrates for graphene growth. Finally, we demonstrate a proof of concept for the merits of ALD based graphene growth for future interconnects (IC) in the form of low temperature patterned PEALD of $MoC_{x}$, for selective CVD growth of MLG, so that no resist residue is present between the graphene and the catalytic material interface. Additionally, we show the advantage of the ALD soft deposition on 100nm thick suspended SiN membrane. Thereafter, MLG was grown on the suspended SiN membranes to achieve suspended graphene based heterostructure without exposing the suspended MLG to wet chemicals or corrosive acids.

\section{Experimental methods}
$MoC_{x}$ thin films have been deposited by plasma enhanced atomic layer deposition (PEALD)  at various temperatures and plasma conditions, as described elsewhere \cite{grady2019tailored}.

\begin{figure}[t!]

	\centering
		\begin{subfigure}[b]{0.7\textwidth}
	\centering
		\includegraphics[width=\linewidth]{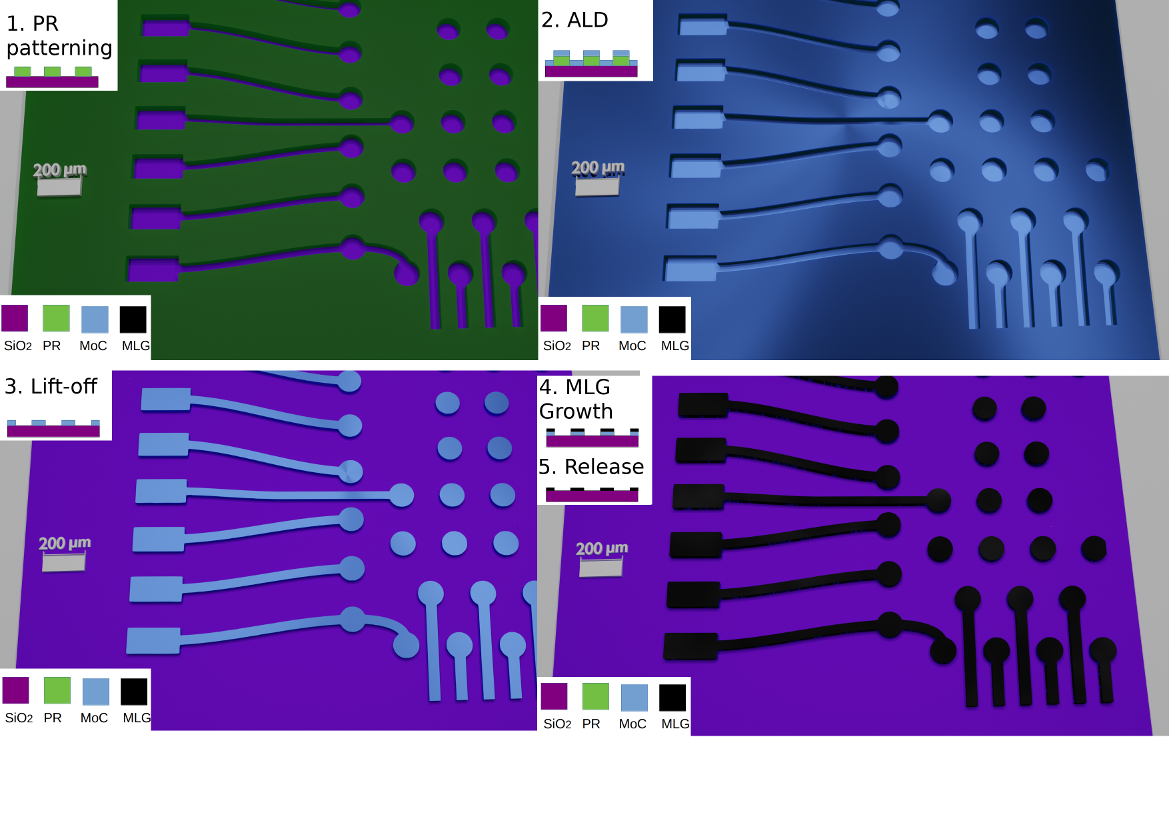}
	\caption{Fabrication flow graphics}
	\label{fig:flow-array-full-portrait-low-r}
\end{subfigure}%
\hfil
		\begin{subfigure}[b]{0.3\textwidth}
		\centering
		\includegraphics[width=\textwidth]{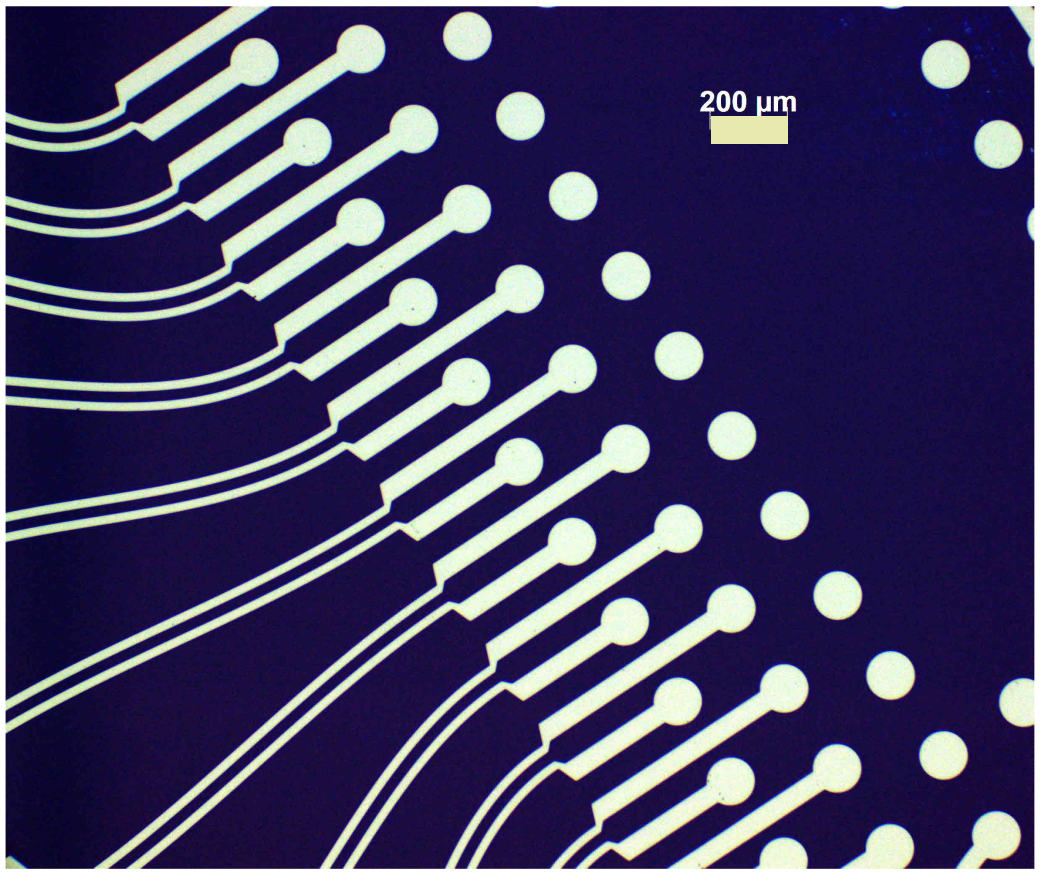}
		\vspace{1.5cm}
		\caption{Optical microscope}
		\label{patternOptic}
	\end{subfigure}%


	\caption{Fabrication of patterned MLG. (a) Fabrication flow schematics: computer grphics and illustration of fabrication steps. (b) Top right: optical microscopy of pre patterned $MoC_{x}$ ALD at 50$^{\circ}C$ after deposition. } 
	\label{fig:fabrication-array}
\end{figure}

\begin{figure}[t!]
	\centering
	\includegraphics[width=\linewidth]{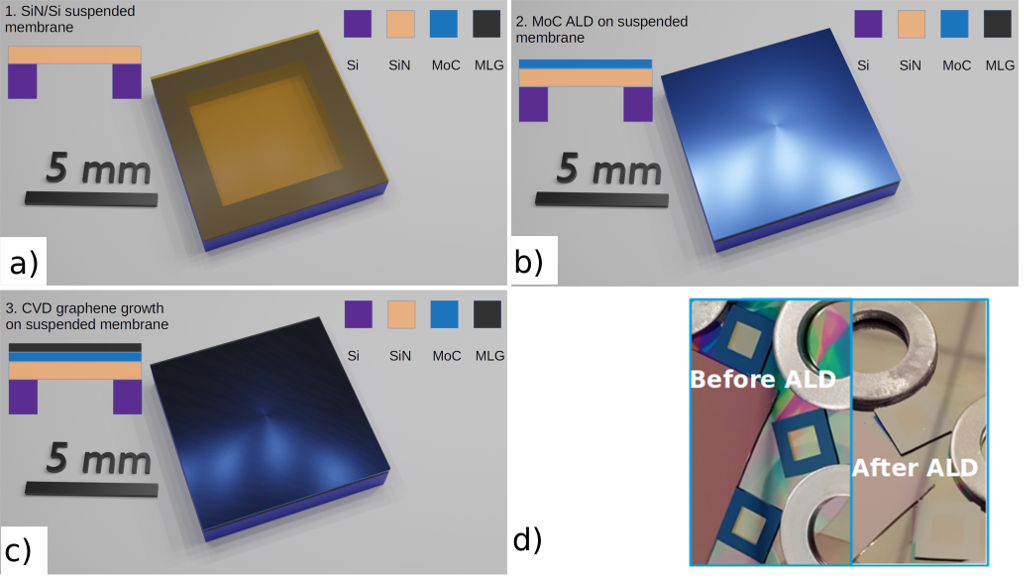}
	
	\caption{Illustration of suspended heterostructure and optical images. (a)+(b)+(c) Computer graphics of the fabrication schematics of suspended graphene heterostructure. (d) Optical images of SiN membranes before and after PEALD of $\sim{15}$ nm $MoC_{x}$ film on top}
	\label{fig:MLG-SiN-Flow}
	
\end{figure}

PEALD was performed on 100 mm Si (100) wafers coated with 450 nm of thermally grown $SiO_2$.  The depositions were performed in an Oxford instruments FlexAL2 ALD reactor, which is equipped with an inductively coupled remote RF plasma (ICP) source (13.56 MHz) with alumina dielectric tube. 
$MoC_{x}$ thin films have been deposited by PEALD  at various temperatures and plasma conditions, with $MoC_{x}$ films varying from 15$\mu{m}$ to 30$\mu{m}$ in thickness.
MLG was grown by low-pressure CVD (LPCVD) in a quartz tube (d=50mm, l=60cm) furnace with 3 heat zones set to 1050$^\circ$C. The typical base pressure when evacuated is $10^{-3}$ mbar. The furnace is set on cart wheels, to allow samples to be rapid annealed, as furnace temperature stabilises within 3.5 minutes after tube insertion. When moved away from the furnace, sample cooling down duration is typically 15 minutes.
Carbon feedstock gas ($CH_4$) is fed along with Argon through a quartz inner tube of 5 mm in diameter to the sealed side of the outer tube.
$MoC_{x}$ films have been saturated with carbon by annealing at temperatures between 500$^{\circ}C$ to 800$^{\circ}C$ with  100 sccm $CH_4$ gas flow at 4 mbar pressure. Then, graphene films  have been grown under similar conditions at 1100$^\circ$C for 10 minutes.
The samples were then promptly extracted from the furnace and allowed to cool down at ambient room temperature under Ar gas flow in the quartz tube.
As shown in figure \ref{fig:flow-array-full-portrait-low-r} photoresist (PR) ma-N 400 with 4.1 $\mu{m}$ thickness on 90 nm $SiO_2$ on Si 2" wafers were used for low temperature PEALD of $MoC_{x}$ film. After deposition, $MoC_{x}$ was patterned by lift-off process, and rinsed in isopropyl alcohol (IPA). 
100 nm and 50 nm SiN membranes were supplied by Philips Innovation Services (PInS) foundry. $MoC_{x}$ were also deposited at 300$^{\circ}C$ on 100 nm SiN membranes suspended on Si (5x5 mm suspended rectangular area).
MLG were then grown the $MoC_{x}$ films as illustrated in figure \ref{fig:MLG-SiN-Flow}. While deposition was successful for both SiN sample thicknesses, due to the brittle nature of the thinner membranes we present here results measured on the thicker 100nm based membranes.
Raman spectroscopy was performed with Reinshaw InVia 514 nm laser.
Film crystallinity and preferred crystal orientation was studied by Gonio x-ray diffraction. Experiments were conducted with PanAnlytical X'pert PROMRD diffractometer operated using $Cu K\alpha (\lambda=1.54A) $.

\section{Characterisation and Results}
This section is divided to three parts: first part deals with MLG growth mechanism in $MoC_{x}$ films and comparison between amorphous and crystalline catalytic substrate. The second part demonstrate a technique for resist and transfer free patterned graphene device. In the third part we fabricate MLG on a thin suspended membrane using ALD and CVD. We show the limitation of Raman measurements on suspended membranes low thermal conductivity. 

\subsection{Carburisation of $MoC_{x}$}
\begin{figure}[t!]
	\centering
	\begin{subfigure}[t]{0.5\textwidth}
		\centering
		\includegraphics[width=\textwidth]{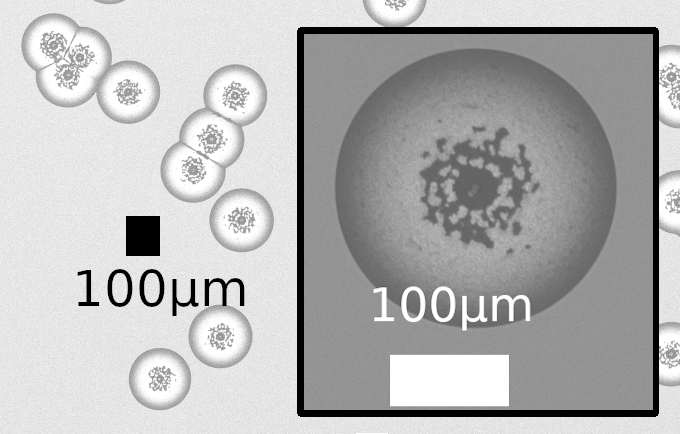}
		\caption{SEM: $MoC_{x}$ ablation}
		\label{SEM:NoMLG1}
	\end{subfigure}%
	\hfill
	
	\begin{subfigure}[t]{0.5\textwidth}
		\centering
		\includegraphics[width=\textwidth]{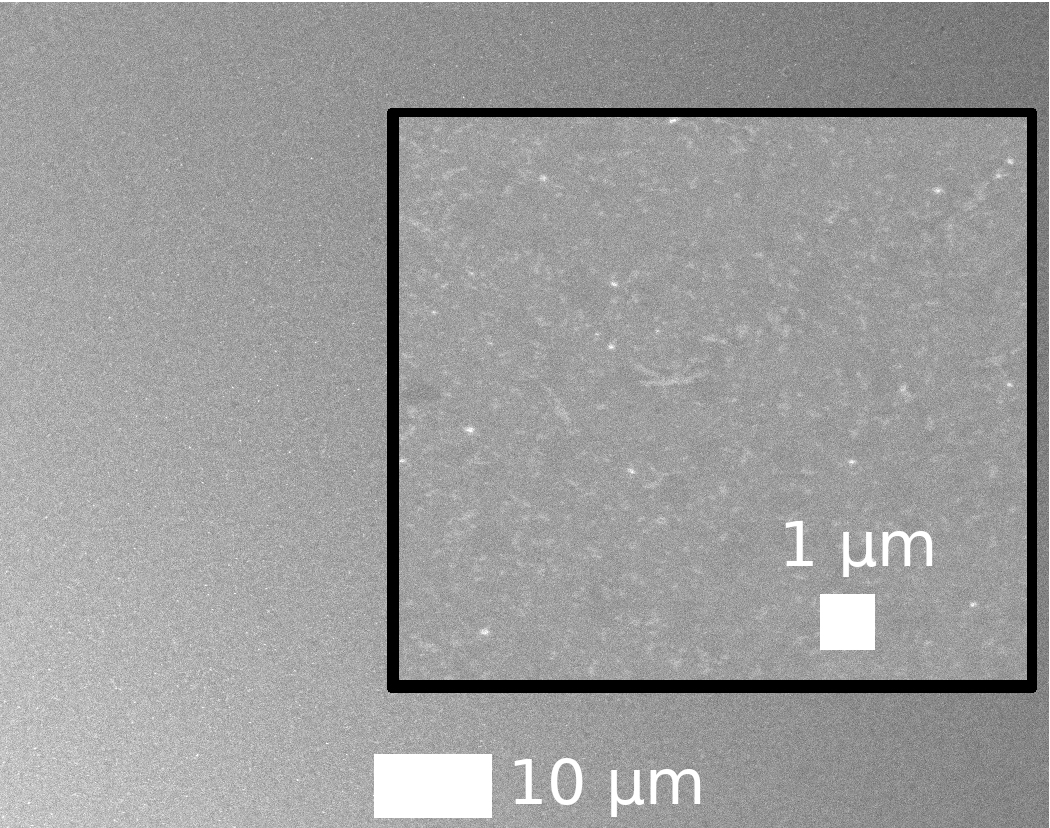}
		\caption{SEM: MLG grown on $MoC_{x}$}
		\label{SEM:MLG}
	\end{subfigure}%
	\hfill
	\begin{subfigure}[t]{0.5\textwidth}
		\centering
	\includegraphics[width=\linewidth]{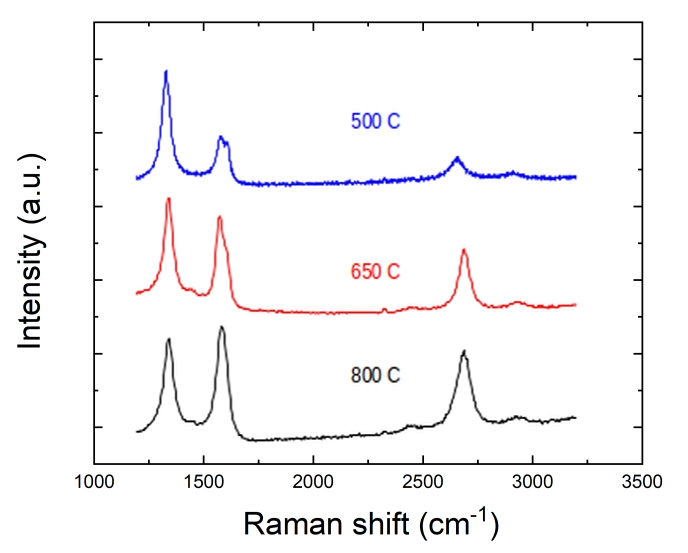}
		\caption{Raman spectra (carb. temp)}
		\label{carburisation_raman}
	\end{subfigure}%
	\caption{SEM images and Raman spectrum of $MoC_{x}$ after $CH_4$ annealing at 1100$^{\circ}C$. Top: (a) SEM image of films that were not carburised prior to growth. Bottom: (b) SEM image of MLG grown on carburised $MoC_{x}$ film. Smooth surface no signs of inhomogeneity. (c) Raman spectrum after graphene growth on crystalline $MoC_{x}$, as a function of carburisation temperature. Growth time remained identical for all 3 experiments.}
	\label{fig:Carburisation}
	
\end{figure}

\begin{figure}[t!]
	\centering
	\includegraphics[width=\linewidth]{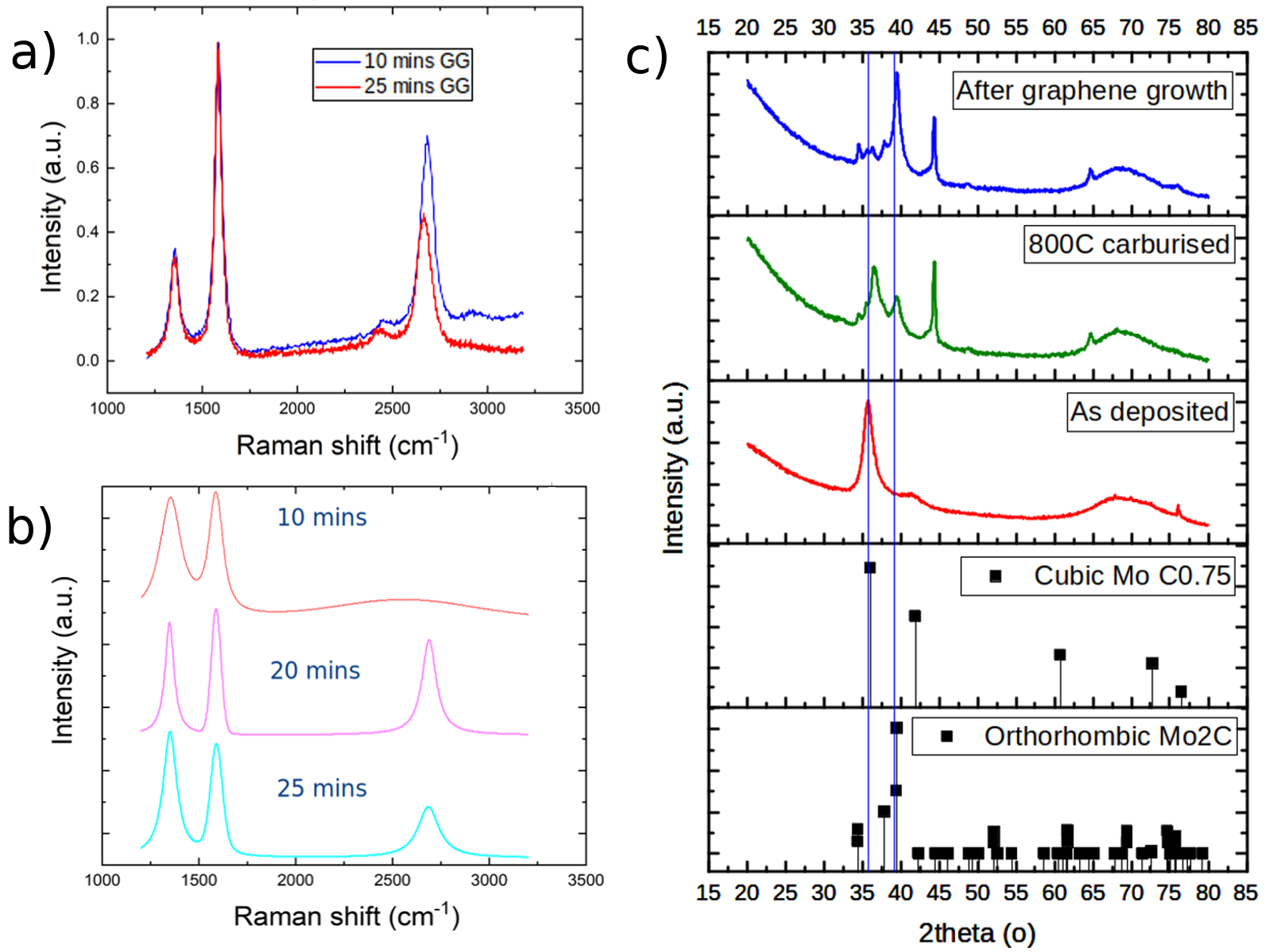}
	\caption{Raman spectra and XRD measurements. Left: Raman measurements of MLG grown on (a) amorphous $MoC_{x}$ film and (b) Single crystalline $MoC_{x}$ after 10 and 25 minutes growth at 1100$^{\circ}C$. Left: (c) XRD measurements of crystalline $MoC_{x}$ for each process step from ALD, to carburisation, to graphene growth}
	\label{fig:Growth_time}
	
\end{figure}

We studied the growth mechanism of graphene on $MoC_{x}$ substrates, to optimise growth process for the various film compositions. The importance of saturating the catalytic $MoC_{x}$ with free carbon is demonstrated and the effects of film crystallinity on carbon precipitation to the surface during growth is shown by Raman and XRD measurements. 
The mechanism of graphene growth on $MoC_{x}$ films is explained in this section. When we subjected the $MoC_{x}$ film to direct growth at 1100$^{\circ}C$ for 10 minutes. SEM images show ablation on the film surface, that has a characteristic graphene Raman signature [\ref{SEM:NoMLG1}]. Outside these areas, no indication of graphene growth was measured. We added then a carburisation step in order to saturate the film with carbon and examined different temperatures around the crystalline phase change temperature ($\sim{650}^{\circ}C$). Figure \ref{carburisation_raman} indicates that for crystalline $MoC_{x}$ film, ideal saturation takes place above the crystalline phase change, for MLG grown at the same growth time. The lack of sufficient carburisation affects the graphene growth significantly, as can be seen in figure \ref{carburisation_raman}. This effect is less dominant with amorphous, low mass density $MoC_{x}$ films, which can be carburised at lower temperatures as well. After establishing an optimal carburisation temperature, we examined the ideal growth time for various $MoC_{x}$ types. As seen in figure \ref{fig:Growth_time}, $MoC_{x}$ with rich carbon content, typically low mass density and crystallinity exhibit good quality MLG growth after 10 minutes at 1100$^{\circ}C$. $MoC_{x}$ films with higher mass density and crystallinity display no graphene growth at this time duration, but rather require a longer exposure time to $CH_4$ at the growth temperature of 20 minutes. Longer exposure begin to deteriorate the graphene film. We study the physical alteration for this crystalline film during the carburisation and graphene growth process, as seen in figure \ref{fig:Growth_time}. XRD diffraction peak typical to cubic- $MoC_{0.75}$ are dominant for the deposited film. After 2 hours carburisation at 800$^{\circ}C$, a transition Orthorhombic crystalline phase is noted, along with sharp graphite (101) plane diffraction peak. After 10 minutes graphene growth at 1100$^{\circ}C$, the orthorhombic phase crystallinity increases, while no significant change in the graphitic peak is noticed. 
\subsection{Patterned MLG Growth}

\begin{figure}[t!]
	\begin{subfigure}{0.5\textwidth}
	\centering
	\includegraphics[width=\textwidth]{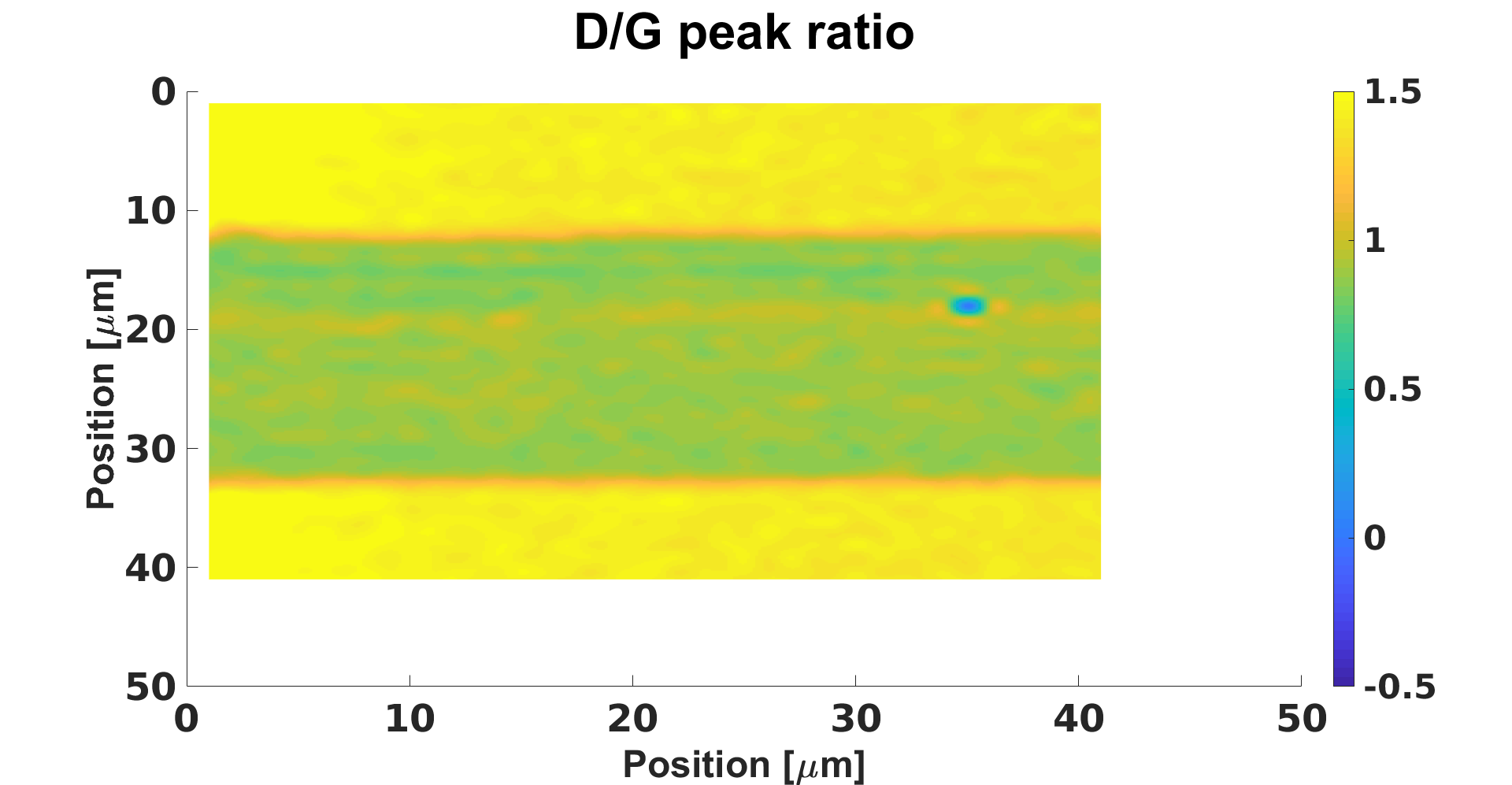}
	\caption{D/G peak ratios mapping}
	\label{patternD-G}
\end{subfigure}%
\hfil
\begin{subfigure}{0.5\textwidth}
	\centering
	\includegraphics[width=\textwidth]{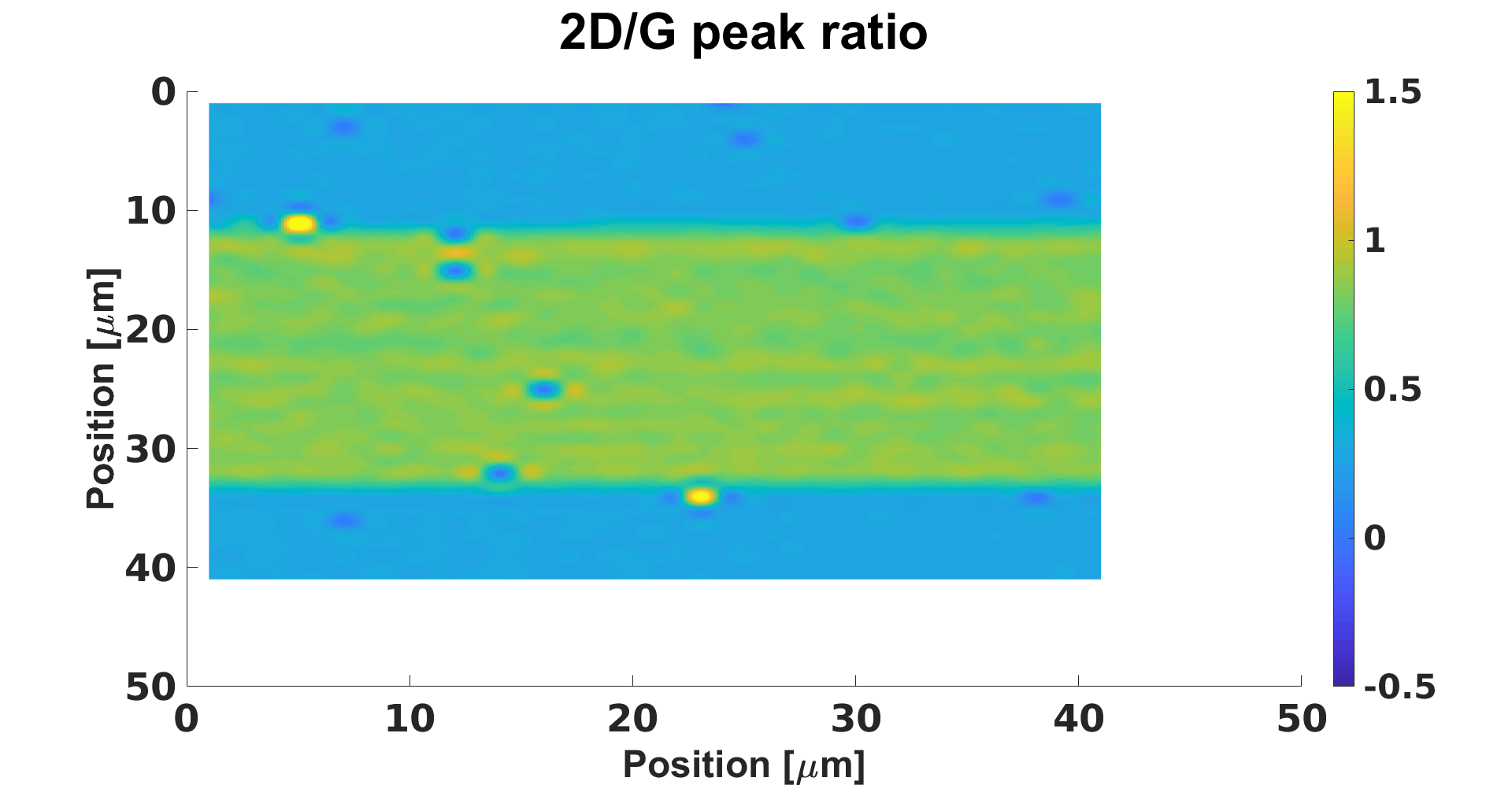}
	\caption{2D/G peak ratios mapping}
	\label{pattern2D-G}
\end{subfigure}%
\caption{Characterisation of patterned MLG: Raman mapping scan of 40$\mu{m}$x40$\mu{m}$ area with 1$\mu{m}$ step resolution of MLG grown on 20 $\mu{m}$ thick $MoC_{x}$ ALD film . (a) left: D/G peak ratio shows a high D/G peak ratio, albeit uniformal continuous coverage. (b) Right: 2D/G peak ratio show a uniform continuous MLG film. Variation in colour is inversely proportional to MLG uniformity} 
\label{fig:Raman map}
\end{figure}

Graphene film has been grown on patterned catalytic substrates. Patterning of $MoC_{x}$ has been performed by a lift-off process, so that no exposed $MoC_{x}$ surface needed to be coated with PR. 
When depositing $MoC_{x}$ at 150$^{\circ}C$ we found resist residues on the $SiO_2$, due to hard baking of the photoresist at this temperature, as the thermal stability limits of the PR (110$^{\circ}C$). Moreover, after the CVD growth we found random patches of MLG coverage on the exposed $MoC_{x}$, and no continuous MLG coverage. However, when the PEALD is performed at 50$^{\circ}C$, we could seamlessly remove the PR and no significant traces were found after lift-off. The CVD growth of graphene on these samples showed full MLG coverage with excellent uniformity, albeit a relatively high D/G peak ratio. Moreover, as can be seen in figure \ref{patternD-G}, D and G peaks were detected throughout the exposed $SiO_2$ areas, but no significant 2D peak.
We have seen that a transfer free release of MLG on $MoC_{x}$ membranes is a direct result of wet etching the Mo based catalyst. This observation is valid for ultra thin layers with thicknesses below 30 nm. Thicker layers ($>50nm$) release the MLG film such that the graphene membrane is afloat on the liquid surface.

\subsection{MLG Growth on Suspended Thin Film}
\begin{figure}[t!]
	\centering
	\begin{subfigure}[b]{0.5\textwidth}
		\centering
	\includegraphics[width=\textwidth]{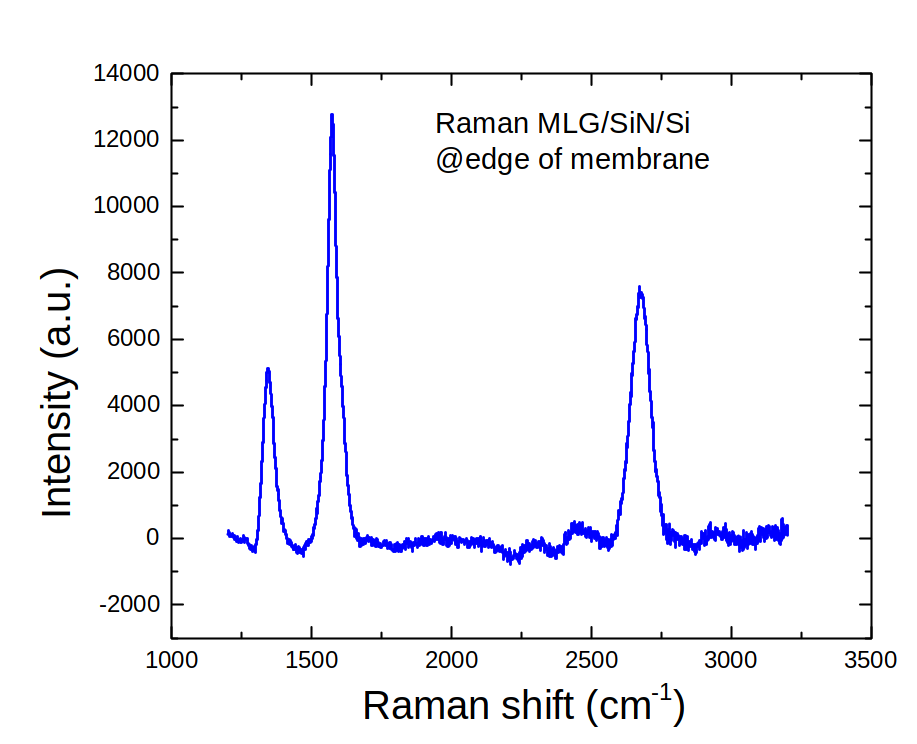}
		\caption{MLG/SiN/Si substrate}
		\label{edge}
	\end{subfigure}%
\hfill
	\begin{subfigure}[b]{0.5\textwidth}
		\centering
		\includegraphics[width=\textwidth]{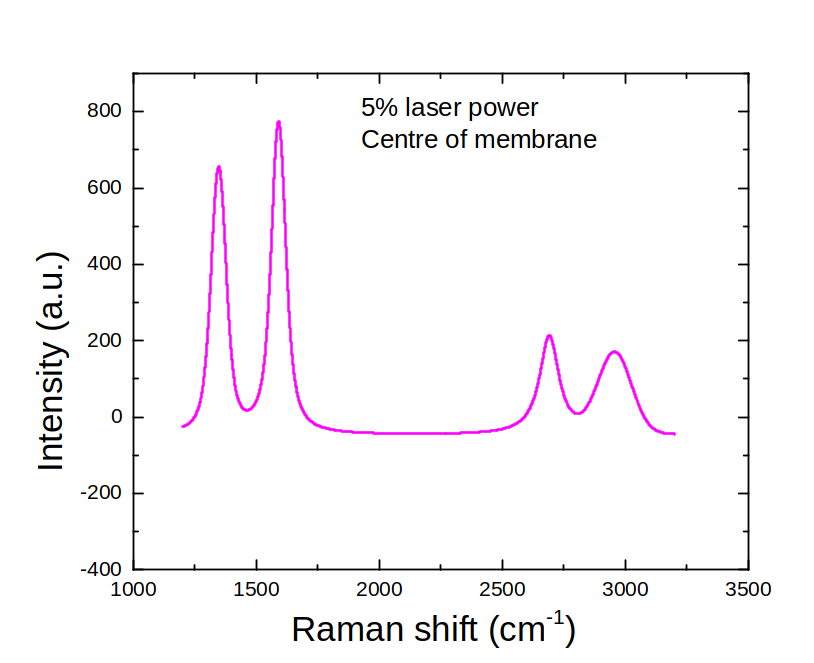}
	\caption{5\% laser power}
		\label{5P}
	\end{subfigure}%
\newline
	\begin{subfigure}[b]{0.5\textwidth}
	\centering
		\includegraphics[width=\textwidth]{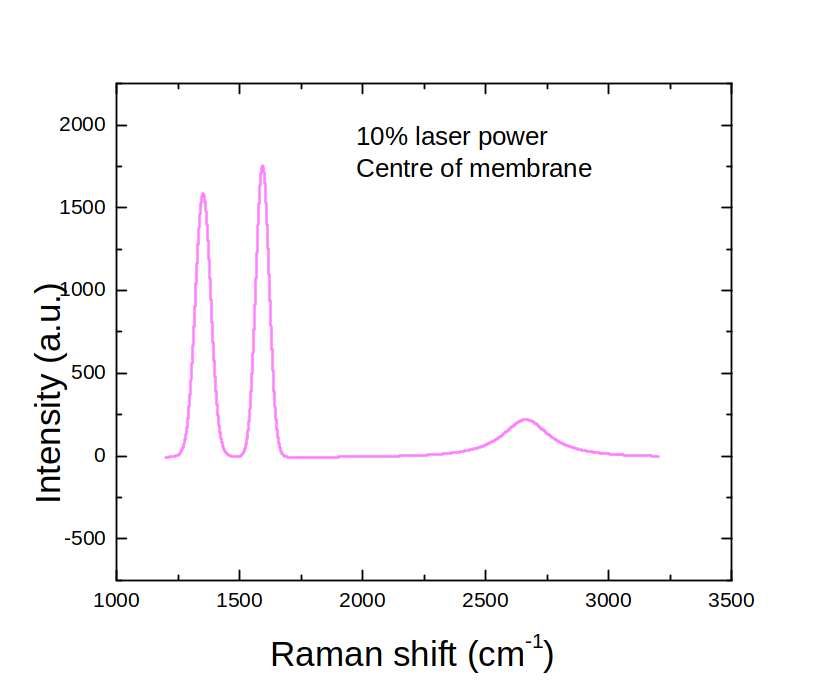}
	\caption{10\% laser power}
	\label{10P}
\end{subfigure}%
\hfill
\begin{subfigure}[b]{0.5\textwidth}
	\centering
		\includegraphics[width=0.5\textwidth]{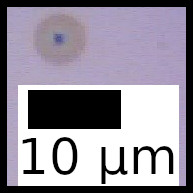}
		\vspace{1cm}
		\caption{Optical microscope: 50\% laser power}
	\label{50P}
\end{subfigure}%
	\caption{Raman spectrum of MLG CVD grown on ALD of $MoC_{x}$ on  SiN membrane. Measurements taken on the membrane's edge with underlying Si, and at various spots around the centre of the membrane with increasing laser power. (a) MLG on SiN membrane's edge with underlying Si substrate. (b) Raman scan measured with 5\% laser power shows diminished 2D peak and a rise in D peak. (c) Raman scan measured with 10\% laser power shows quenched 2D peak and a high D peak. (d) Optical microscope image: after Raman scan measured with 50\% laser power. A hole is punched through the MLG/MoC/SiN suspended heterostructure}
	\label{fig:suspended_SiN}
	
\end{figure}

SiN membranes were suspended on Si substrate with 5x5 mm openings. $MoC_{x}$ was then deposited by PEALD process. There after, MLG was grown by a CVD growth process. Thus, a suspended heterostructure of MLG/$MoC_{x}$/SiN was fabricated on the supporting Si frame.
MLG film grown on SiN membrane were characterised by Raman spectroscopy, as can be seen in \ref{fig:suspended_SiN}. We measured the Raman signal at the edge of the suspended membrane where the SiN was supported by the underlying Si, and at the centre of the membrane, where the MLG was on top of $\sim{15}$nm $MoC_{x}$ and 100 nm SiN.
 The Raman spectrum was fitted and baseline corrected due to enhanced SiN background signal. Although Raman spectroscopy is considered a non-destructive measurement,  we discovered that suspended MLG/SiN membranes were highly sensitive to the Raman laser power. As figure \ref{10P} shows, when 10\% power of the 20mW Laser was used, an increase in the D peak was measured, and the 2D peak was quenched in comparison to 5\% power. With 50\% power of the Raman's laser, we could punch a hole through the heterostructure suspended membrane as seen with an optical microscopy image [see figure \ref{50P} ].
 
\section{Discussion and Conclusions}
We have demonstrated in this work the growth mechanism and conditions of MLG on PEALD $MoC_{x}$ films. In order to achieve full graphene film coverage, and a uniform graphene growth, carburisation of the substrate is essential. The carburisation step allows for saturation of the catalytic film with free carbon. The free carbon then precipitate to the surface during annealing at 1100$^\circ$C. We found that in order to saturate $MoC_{x}$ film with higher mass density, carburisation temperature has to exceed the crystalline phase change point. For $MoC_{x}$ film we found that point to be around $\sim{650}$. Amorphous films with lower mass density could be saturated at lower carburisation temperatures. \newline 
We compared MLG growth time on amorphous film and on a highly crystalline one. The prolonged annealing time necessary to grow graphene on crystalline film is stipulated to result from low carbon saturation during carburisation, due to the dense crystal structure and the resulting low bulk precipitation of free carbon. Defects ratio of MLG on crystalline film could be explained with the transition from a single crystalline cubic-$MoC_{0.75}$ to a polycrystalline orthorhombic phase, which results in multiple strain points in the MLG film. By contrast, graphene growth on an amorphous carbon rich film is more facile and results in a lower D/G peak ratios.
After understanding the growth mechanism, we demonstrated the advantages of graphene growth on PEALD catalytic films. The low temperature, soft and atomic precise deposition allows for pre patterning of catalytic substrates, such that no PR coating of the growth surface is needed. We demonstrated here an initial proof of concept, which showed a full film coverage with excellent uniformity. We speculate that decomposition of the  $CH_4$ molecules on the $SiO_2$ surface with PR residue is the reason for the carbide formation outside the patterned $MoC_{x}$ areas. The relatively high D/G peak could be addressed by growth process optimisation, and different approaches to selective ALD which was out of the scope of this work. We cannot rule out potential surface impurities during the lift-off process, that could result in additional amorphous carbon during the CVD growth process. A complete bottom up selective area ALD process will avoid any potential resist contamination and should allow for a broader deposition window.
We have also used ALD to deposit $MoC_{x}$ films on brittle SiN membranes of 50 and 100 nm. The MLG growth demonstrated on 100 nm suspended membrane shows a viable fabrication route for sensor applications, and graphene based resonators for a wide range of frequencies.  We showed for the first time damage to a suspended heterostructure caused directly by Raman measurements. The low thermal conductivity of the SiN is likely the cause for the local damage to the suspended heterostructure, with no directly available heat sink - as oppose to the membrane edges. This however could prove useful, when punctured suspended graphene heterostructures are required, for water filtration and desalination applications for example.
Building on the capabilities demonstrated here, the route for future application based on patterned graphene or suspended graphene heterostructure is clearly marked. Further steps, such as $Al_{2}O_{3}$  encapsulation, could be readily performed by ALD directly on the MLG without damage by a process of hydrogenation and post ALD annealing \cite{vervuurt2017uniform}.
Moreover, one can combine recent developments in area selective ALD to realise a complete bottom up fabrication of the catalytic substrate resist free, with atomic-scale precision alignment.

\subsection*{Acknowledgements}
This  research  is  supported  by  the Dutch  Technology  Foundation STW (project number 140930), which is part of the Netherlands Organization for Scientific Research (NWO), and partly funded by the Ministry of Economic Affairs as well as ASML and ZEISS. E. Grady thanks Cristian Helvoirt, Janneke Zeegbregts, Jeroen van Gerwen and the lab technical staff for their support.
\bibliographystyle{unsrt}
\bibliography{MLG}

\newpage

\bigskip
\section*{Figures}
%
%
%

\newpage

\end{document}